\documentclass[twocolumn, showpacs,amsmath,amssymb]{revtex4}
\usepackage{graphicx}
\usepackage{epstopdf}
\usepackage{bm}
\usepackage{dcolumn}
\usepackage{bm}

\begin{document}

\preprint{APS/123-QED}

\title{Dynamic cancellation of ac Stark shift for pulsed EIT/Raman optical lattice clocks}

\author{Thomas Zanon-Willette, Andrew D. Ludlow, Sebastian Blatt, Martin M. Boyd, Ennio Arimondo,* Jun Ye}
\affiliation{JILA, National Institute of Standards and Technology
and University of Colorado, Department of Physics, University of
Colorado, Boulder, Colorado 80309-0440, USA}
\date{\today}

\begin{abstract}

We propose a combination of Electromagnetically Induced Transparency
(EIT)/Raman and pulsed spectroscopy techniques to accurately cancel
frequency shifts arising from EIT fields in forbidden optical
lattice clock transitions of alkaline earth atoms. Time-separated
laser pulses are designed to trap atoms in coherent superpositions
while eliminating off-resonance ac Stark contributions at particular
laser detunings from the intermediate excited state. The scheme
achieves efficient population transfer up to $60\%$ with potential
inaccuracy $<$ $10^{-17}$. Both complex wave-function formalism and
density matrix approach predict cancellation of external light
shifts at the mHz level of accuracy, under low field strengths or
short interaction times.
\end{abstract}


\maketitle

\begin{figure}[!ht]
\centering%
\resizebox{8 cm}{!}{
\includegraphics[angle=0]{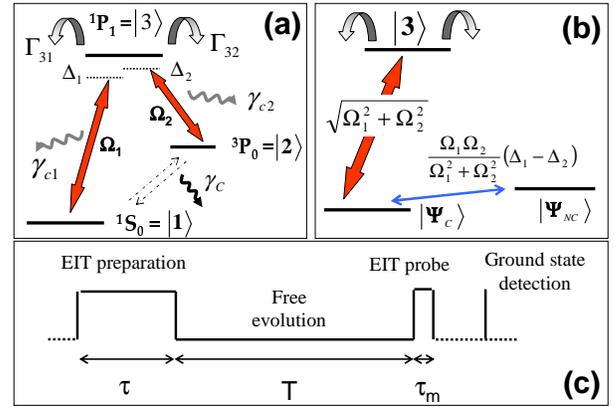}}
\caption{(Color online)(a) Three level atom-light configuration for
an optical lattice clock based on time-separated laser pulses
including relaxation and decoherence rates. The optical detunings
$\Delta_{1}\equiv\Delta_{0}+\eta_1$,
$\Delta_{2}\equiv\Delta_{0}-\delta_{r}+\eta_2$ include ac
Stark-shifts $\eta_i$ from off-resonant levels. Here $\Delta_{0}$ is
the common mode detuning and $\delta_r$ denotes deviation from the
Raman condition. (b) The corresponding dressed-state representation
of bright $|\Psi_C\rangle$ and dark $|\Psi_{NC}\rangle$ states
defining the clock transition. (c) The probing pulse
sequence.}\label{scheme-lambda}
\end{figure}

In the field of optical frequency standards and clocks, single
trapped ions \cite{ion} and alkaline earth atoms
\cite{neutral,Takamoto:2005,Ludlow:2006,Le Targat:2006,Barber:2006}
are advancing clock performances. The advantage arises from
superhigh resonance quality factors of these optical transitions
\cite{Boyd:2006}, which are expected to be $10^{5}$ better than
microwave fountains. These fountain clocks are already below the
$10^{-15}$ relative fractional uncertainty \cite{Bize:2005}.
Fermionic isotopes of alkaline earths trapped in optical lattices at
the magic wavelength \cite{Katori:2003} offer ultra narrow
linewidths of a few mHz without recoil and Doppler effects, but
remain potentially sensitive to systematic effects arising from the
nuclear spin-related tensor polarizability
\cite{Takamoto:2005,Ludlow:2006,Le Targat:2006}. On the other hand,
bosonic isotopes with no nuclear spin and a higher natural isotopic
abundance avoid multiple hyperfine components but lack direct
excitation of the clock transition $|1\rangle \leftrightarrow
|2\rangle$ in Fig.~\ref{scheme-lambda}(a). Indirect excitation via
continuous-wave Electromagnetically Induced Transparency (EIT) has
been proposed to probe these forbidden transitions
\cite{Santra:2005,Hong:2005}. A similar scheme for the $^{174}$Yb
forbidden clock transition was implemented by applying a dc magnetic
field for state mixing \cite{Barber:2006}.

All such schemes can suffer from Stark shifts due to non-resonant
electric-dipole couplings of the clock levels to other states
induced by the applied electromagnetic fields
\cite{Haffner:2003,Sherman:2005}. Ref. \cite{Santra:2005} provides
some detailed calculations of these shifts. To further reduce this
potential systematic error, we could apply an approach similar to
that used for the determination of the magic wavelength
\cite{Takamoto:2005,Ludlow:2006} or the hyperpolarizability
contribution to the ac Stark shifts \cite{Brusch:2006}: Measurements
at different field strengths are used to extrapolate the clock
frequency to vanishing field. However, this simple approach does not
apply to the EIT-related schemes where the applied field strength
modifies also the optical pumping time required to prepare the atoms
in a coherent superposition \cite{Fleischhauer:2005}. The
preparation time required for optimal signal contrast and clock
stability becomes unpractically long at low field strengths. But
using large fields increases the ac Stark shifts and limits the
clock accuracy. To overcome these limits, the pulsed scheme proposed
in this Letter (Fig.~\ref{scheme-lambda}(c)) optimizes clock
performance by utilizing time-separated laser pulses to prepare and
interrogate the optical clock transition \cite{Knight:1982}. It is
an original mix of Ramsey spectroscopy \cite{Morigana:1993} and
highly efficient population transfer under Coherent Population
Trapping (CPT) \cite{Arimondo:1996}. The first pulse prepares atoms
in a coherent superposition and the second pulse probes the clock
frequency. This configuration produces a large contrast in the
detected clock signal. More importantly, as the detunings of the
applied fields affect the phase evolution of the atomic
wave-function, a proper combination of the common mode laser
detuning $\Delta_{0}$ and pulse durations $\tau,\tau_{m}$ (Fig. 1)
reduces the clock shift to $\sim10^{-17}$. The discussion presented
here reveals for the first time a general relation connecting the
preparation time of the Raman coherence and the signal contrast in
the subsequent detection of this coherence, relevant to many EIT or
CPT related experiments.

The atomic evolution between $^{1}S_{0}$ and $^{3}P_{0}$ is properly
described in the dressed state picture
(Fig.~\ref{scheme-lambda}(b)),
\begin{equation}
\small{
\begin{split}
|1\rangle&=\frac{\Omega_{1}}{\sqrt{\Omega_{2}^{2}+\Omega_{1}^{2}}}|\Psi_{C}\rangle+\frac{\Omega_{2}}{\sqrt{\Omega_{2}^{2}+\Omega_{1}^{2}}}|\Psi_{NC}\rangle\\
|2\rangle&=\frac{\Omega_{2}}{\sqrt{\Omega_{2}^{2}+\Omega_{1}^{2}}}|\Psi_{C}\rangle-\frac{\Omega_{1}}{\sqrt{\Omega_{2}^{2}+\Omega_{1}^{2}}}|\Psi_{NC}\rangle\\
\end{split}}\label{superpositions}
\end{equation}
where $\Omega_{1}$ and $\Omega_{2}$ are the Rabi frequencies for the
transitions $^{1}S_{0}\leftrightarrow{^{1}P_{1}}$ and
$^{3}P_{0}\leftrightarrow{^{1}P_{1}}$. For an ideal 3-level system
described in Eq.~(\ref{superpositions}), the dark state
$|\Psi_{NC}\rangle$ remains insensitive to light shift, while the
bright state $|\Psi_{C}\rangle$ is always coupled to the laser
light. A realistic atomic clock has to deal with off-resonant ac
Stark shifts acting on $|\Psi_{C}\rangle$ while atoms are pumped
into $|\Psi_{NC}\rangle$ with a few spontaneous emission cycles.
Thus, a judicious trade-off between the short-time dynamics for a
high-contrast signal (large optical pumping) and the reduced
external ac shifts (and resonance power broadenings) under a low
field strength needs to be found for practical realizations of these
EIT/Raman-type clocks.

To describe our pulsed method, we start from a three-level
configuration as shown in Fig.~\ref{scheme-lambda}(a). The Optical
Bloch Equations (OBEs) describe three-level dynamics including
external shifts, relaxations, and decoherences between atomic states
\cite{Santra:2004} in terms of the density matrix:
\begin{equation}
\small{\dot{\rho}=-\frac{i}{\hbar}[H,\rho]+\mathcal{R}\rho}
\label{pilot-equation}
\end{equation}
In the interaction picture, the atom-light hamiltonian $H$ and
relaxation matrix $\mathcal{R}\rho$ become
\begin{equation}
\small{\frac{H}{\hbar}= \begin{pmatrix}
\Delta_{1}&0&\Omega_{1}\\
0&\Delta_{2}&\Omega_{2}\\
\Omega_{1}&\Omega_{2}&0\\
\end{pmatrix}; \mathcal{R}\rho=
\begin{pmatrix}
\Gamma_{31}\rho_{33}&-\gamma_{c}\rho_{12}&-\gamma_{c1}\rho_{13}\\
-\gamma_{c}\rho_{21}&\Gamma_{32}\rho_{33}&-\gamma_{c2}\rho_{23}\\
-\gamma_{c1}\rho_{31}&-\gamma_{c2}\rho_{32}&-\Gamma\rho_{33}\\
\end{pmatrix}}\label{hamiltonian-matrix}
\end{equation}
The relaxation matrix includes the spontaneous emission rates
$\Gamma=\Gamma_{31}+\Gamma_{32}$, optical decoherences
$\gamma_{c1},\gamma_{c2}$, and the Raman decoherence $\gamma_{c}$
(see Fig.~1(a)). Electric and/or magnetic dipole couplings determine
the Rabi frequencies $\Omega_i$ ($i=1,2$).
Eq.~(\ref{pilot-equation}) describes the dynamics of a closed
$\Lambda$-system where optical detunings $\Delta_i$ include ac Stark
shifts $\eta_i$ from non-resonant electric-dipole couplings of
$|1\rangle$ and $|2\rangle$ to other states. For
$\Omega_{1},\Omega_{2}\lesssim\Gamma_{31},\Gamma_{32},\gamma_{c1},\gamma_{c2}$,
the population in state $|3\rangle$ is slaved to the population
difference $\Delta n(t)\equiv\rho_{22}(t)-\rho_{11}(t)$ and Raman
coherence $\rho_{12}(t)$. This allows finding analytical solutions
to Eq.~(\ref{pilot-equation}) by adiabatic elimination of the
intermediate state $|3\rangle$ \cite{Hemmer:1989,Shahriar:1997}. The
reduced two-level system dynamics are described by a Bloch-vector
representation \cite{Jaynes:1955,Steinfeld:1978}.
\begin{figure}[!t]
\centering%
\resizebox{\linewidth}{!}{
\includegraphics[angle=0]{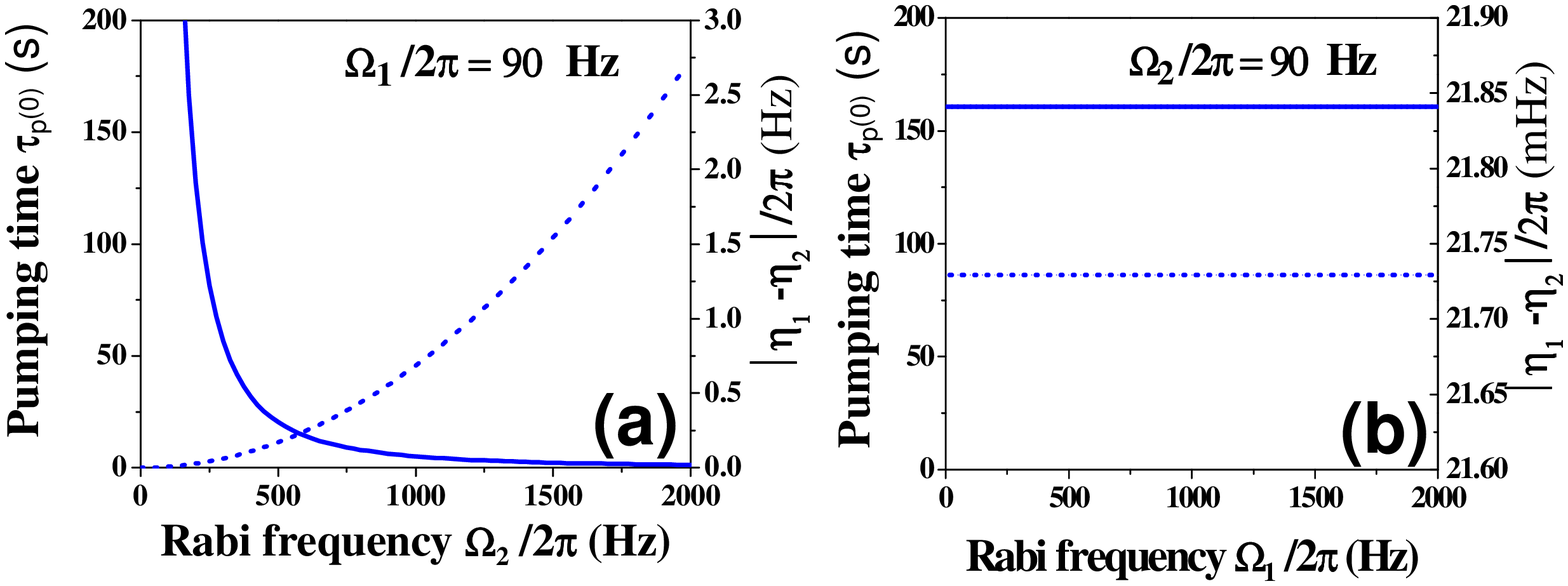}}
\caption{Differential ac Stark shifts $\eta_1-\eta_2$ (dashed
curves) on the $^{1}S_{0}\leftrightarrow {^{3}P_{0}}$ clock
frequency and the optical pumping time $\tau_{p}(\Delta_{0}=0)$
(solid curves) using Eq.~(\ref{time}) vs either (a) magnetic Rabi
frequency $\Omega_{2}$ or (b) electric Rabi frequency
$\Omega_{1}$.}\label{time-versus-shift}
\end{figure}

To remove ac Stark shifts while maintaining a high signal contrast,
we apply the Ramsey technique for EIT/Raman fields to this effective
two-level system, minimizing systematic frequency shifts over the
free-evolution time $T$. The Ramsey-like sequence of preparation,
free-evolution, and probe, followed by the final destructive
detection of the ground state population, is indicated in
Fig.~\ref{scheme-lambda}(c). This eliminates power broadening of the
clock transition which is always present for continuous excitation
\cite{Zanon:2005}. By solving for the two-level system using the
methods in \cite{Steinfeld:1978} we can express the populations as

\begin{equation}
\small{{\rho_{ii}\equiv\alpha_{ii}(\tau,\tau_{m})\left(1+\beta_{ii}(\tau,\tau_{m})
e^{-\gamma_{c}T}\cos[\delta_{r}T-\Phi(\tau,\tau_{m})]\right)}}
\label{pulsed-two-photon}
\end{equation}
where $\alpha_{ii}(\tau,\tau_{m})$ is the overall envelope function
and $\beta_{ii}(\tau,\tau_{m})$ is the amplitude of fringes, both
containing exponential decays $e^{-\tau/\tau_{p}}$ and
$e^{-\tau_{m}/\tau_{p}}$ to their steady states \cite{Jaynes:1955}.
$\tau_p$ is the characteristic optical pumping time. The atomic
phase shift $\Phi$ produces an approximated clock frequency shift
assuming $\tau,\tau_m\lesssim T$:
\begin{equation}
\small{
\begin{split}
\delta\nu=\frac{\Phi(\tau,\tau_{m})}{2\pi T(1 +
\frac{\tau+\tau_m}{2T})},
\end{split}}\label{EIT-phase-shift}
\end{equation}
which includes all ac Stark contributions accumulated during the
pulsed interactions. Hence, a longer free-evolution time $T$ reduces
the light shifts on the clock transition. Furthermore, as will be
shown below, a special value $(\Delta_0)_m$ of the common detuning
$\Delta_0$ can be found to suppress ac Stark effects on the clock
frequency. Study of the population dynamics from
Eq.~(\ref{pulsed-two-photon}) leads to an expression for the time
$\tau_{p}$ that is required to pump atoms into their final steady
state, simplified for $\Delta_{0}\simeq\Delta_{1}\simeq\Delta_{2}$:
\begin{equation}
\small{
\tau_{p}(\Delta_{0})\approx\frac{2}{\Gamma}\frac{\Delta_{0}^{2}+\Gamma^{2}/4}{(\Omega_{1}^{2}+\Omega_{2}^{2})}
\left[1-\Upsilon\left(\frac{\Omega_{1}^{2}-\Omega_{2}^{2}}{\Omega_{1}^{2}+\Omega_{2}^{2}}\right)\right]^{-1}}.
\label{time}
\end{equation}
Here $\Upsilon=\left(\Gamma_{31}-\Gamma_{32}\right)/\Gamma$ is the
branching ratio difference for the intermediate state which scales
the contribution of each Rabi frequency to the pumping rate
$\tau_{p}^{-1}$. We emphasize the importance of this time scale as
it determines experimental protocols for detecting the EIT or CPT
response in either transient or steady states. Previous work on EIT
or CPT concentrates mainly on the symmetric case with $\Upsilon=0$.
But in the case of alkaline earths where $\Upsilon\sim\pm1$,
Eq.~(\ref{time}) shows that the Rabi frequency associated with the
weaker transition dictates $\tau_{p}$. For the $^{88}$Sr lattice
clock where $\Gamma_{31}=2 \pi \times 32$ MHz $\gg\Gamma_{32}= 2\pi
\times 620$~Hz (i.e. $\Upsilon\sim1$), the pumping time at resonance
$\tau_{p}(0)$ is determined by the magnetic dipole coupling
$\Omega_{2}$ between $^{3}P_{0}$ and $^{1}P_{1}$.
Figure~\ref{time-versus-shift} shows the dependence of $\tau_{p}(0)$
on each Rabi frequency while keeping the other one fixed. The dotted
lines are the corresponding differential ac Stark shift of the clock
frequency in the steady state regime. Note that small ac Stark
shifts correspond to long optical pumping times conflicting with
realistic clock duty cycles. For instance, the proposal by
\cite{Santra:2005} with ac Stark shift below 21.7~mHz for an
accuracy of $2\times10^{-17}$ leads to a signal contrast of a few
$\%$ only after 160~s. The scheme presented here finds a combination
of parameters that maximizes contrast while suppressing ac Stark
shifts, exploiting the transient dynamics for short pulses and
detuned laser fields. Note that due to the highly asymmetric
$\Upsilon$, this scheme can uniquely exploit ground-state detection
with a high-contrast narrow resonance manifested in the atomic
population transfer \cite{Jyotsna:1995}. In the region of detuning
between Raman spectroscopy ($\Delta_0/\Gamma\gg 1$) and EIT/CPT
($\Delta_0/\Gamma\ll 1$), we find contrasts of up to 60\%, even
though $\tau\ll\tau_p(\Delta_0)\simeq 100$~s. This same approach
could be extended easily to the four level scheme \cite{Hong:2005},
the magnetic induced optical transition \cite{Barber:2006}(with the
magnetic field and the common detuning as operational parameters),
or any other clock configurations involving dark states.
\begin{figure}[!t]
\centering \resizebox{\linewidth}{!}{
\includegraphics[angle=0]{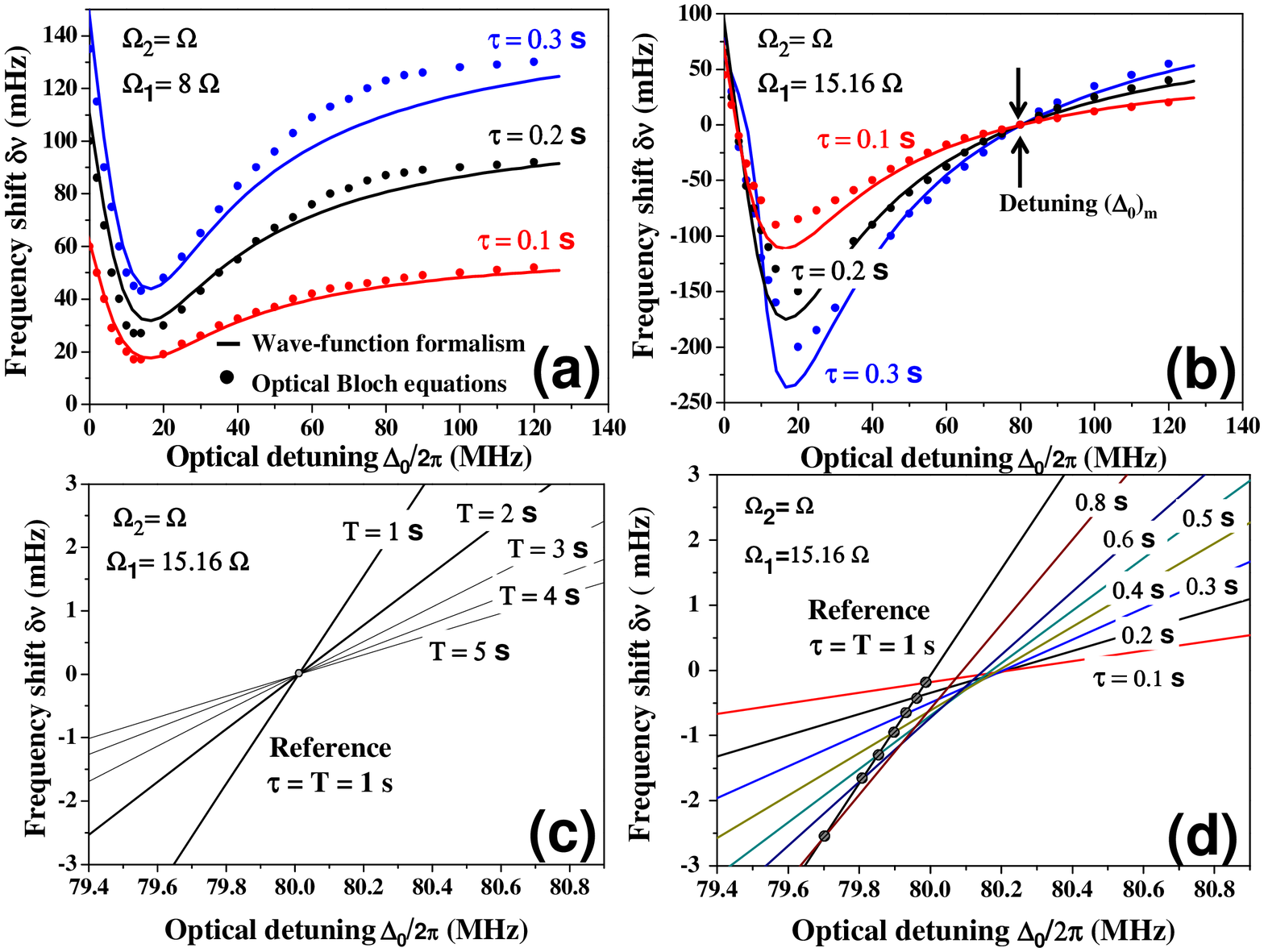}}
\caption{(Color online)Time diluted frequency shift
(Eq.~(\ref{EIT-phase-shift}) and
Eq.~(\ref{ground-state-phase-shift})) arising from off-resonance ac
Stark shift contributions to the
$^{1}S_{0}\leftrightarrow{^{3}P_{0}}$ transition under different
optical detunings $\Delta_{0}$. (a) Three different cases of pulse
durations $\tau=\tau_{m}$ are shown, under $T=1$~s and
$\Omega_1/\Omega_2$ = 8. Numerical calculations based on
Eq.~(\ref{pilot-equation}) (solid dots) agree with analytical
results from the wave-function formalism. The pumping time at
resonance is $\tau_{p}(0)=5$~s and the common Rabi frequency is
$\Omega=\sqrt{\Gamma/4\tau_{p}(0)}$. (b) Same as (a) except
$\Omega_1/\Omega_2$ = 15.16, showing Stark shift cancelation near
$(\Delta_{0})_m=80~$MHz. (c) A zoomed-in plot of $\delta\nu$ versus
$\Delta_0$, with the slope reduced for a longer $T$. The location of
the common crossing point is $(\Delta_0)_m$. (d) $\delta\nu$ under
different $\tau$ around $(\Delta_{0})_m$. The crossings (shown as
solid dots) between lines indicate that the same $\delta\nu$ is
obtained for different values of $\tau$.
}\label{magic-optical-shift}
\end{figure}

The small difference between the field-free clock detuning
$\delta_{r}$ and the ac Stark shifted detuning
$\Delta_{1}-\Delta_{2}=(\eta_1-\eta_2)+\delta_{r}$ under laser
fields leads to a small phase shift of the Ramsey-EIT fringe defined
by $\Phi(\tau,\tau_{m})$ in Eq.~(\ref{EIT-phase-shift})
\cite{Hemmer:1989,Shahriar:1997}. Solving Eq.~(\ref{pilot-equation})
numerically, we find that a judicious choice of the laser detuning
$(\Delta_0)_m$ cancels the external ac Stark shifts, minimizing the
influence to the clock transition when high field strengths are used
to rapidly drive EIT resonances. To confirm these results, we also
establish an analytical expression for $\Phi(\tau,\tau_{m})$ based
on the atomic wave-function formalism \cite{Dalibard:1992}, using
the Hamiltonian of Eq.~(\ref{hamiltonian-matrix}) adding only the
term $-i\Gamma/2$ associated with spontaneous relaxation
\cite{Radmore:1982,Stalgies:1998} and neglecting all lattice
decoherences. By adiabatic elimination of state $|3\rangle$, within
an effective two-level system including only the clock states
$|1\rangle$ and $|2\rangle$, the amplitudes evolve with a matrix
$M$, generalized from Ref.\cite{Zanon:2005a} by assuming $\Delta_1
\neq \Delta_2$:
\begin{equation}
\small{
\begin{split}
M &= \begin{pmatrix} \cos\left(\frac{\omega}{2}t\right)+
i\frac{\Delta_\text{eff}}{\omega}\sin\left(\frac{\omega}{2}t\right)
&
2i\frac{\Omega_\text{eff}}{\omega}\sin\left(\frac{\omega}{2}t\right) \\
2i\frac{\Omega_\text{eff}}{\omega}\sin\left(\frac{\omega}{2}t\right)
& \cos\left(\frac{\omega}{2}t\right)-
i\frac{\Delta_\text{eff}}{\omega}\sin\left(\frac{\omega}{2}t\right)\end{pmatrix}\\
&\equiv \begin{pmatrix} M_{+} & M_{\dagger} \\ M_{\dagger} &
M_{-}\end{pmatrix}
\end{split}}
\end{equation}
where
$\omega=\left(\Delta_\text{eff}^{2}+4\Omega_\text{eff}^{2}\right)^{1/2}$
and $\Delta_\text{eff}$ ($\Omega_\text{eff}$) is the complex
detuning (Rabi frequency) in the effective two-level system,
extending the definitions of \cite{Moler:1992}:
\begin{equation}
\small{
\begin{split}
\Delta_\text{eff}&=\Omega_{1}^{2}\frac{\Delta_{1}+i\Gamma/2}{\Delta_{1}^{2}+\Gamma^{2}/4}-\Omega_{2}^{2}\frac{\Delta_{2}+i\Gamma/2}{\Delta_{2}^{2}+\Gamma^{2}/4}-(\Delta_{1}-\Delta_{2})\\
\Omega_\text{eff}&=\Omega_{1}\Omega_{2}\left(\frac{\Delta_{1}+i\Gamma/2}{\Delta_{1}^{2}+\Gamma^{2}/4}\times\frac{\Delta_{2}+i\Gamma/2}{\Delta_{2}^{2}+\Gamma^{2}/4}\right)^{1/2}
\end{split}}
\end{equation}
The atomic phase depends not only on the wave-function coefficients
of the atomic evolution but also on the steady states included in
the closed density matrix equations \cite{Orriols:1979}. However,
when short pulses $\tau,\tau_{m}\ll\tau_{p}(\Delta_0)$ are applied,
stationary solutions can be ignored. For initial conditions
$\rho_{11}(0)=1,\rho_{22}(0)=0$, we find an expression for the
atomic phase related to the clock frequency shift:
\begin{equation}
\small{
\begin{split}
\Phi(\tau,\tau_{m})\sim
\mathrm{Arg}\left[\frac{M_{-}(\tau_{m})}{M_{\dagger}(\tau_{m})}\frac{M_{\dagger}(\tau)}{M_{+}(\tau)}\right]
\end{split}}\label{ground-state-phase-shift}
\end{equation}
We are able to find values of $(\Delta_0)_m$ where the clock shift
is suppressed for different practical choices of Rabi frequencies
$\Omega_i$. Fig.~\ref{magic-optical-shift}(a) plots the clock
frequency shift ($\delta\nu$ as defined in
Eq.~(\ref{EIT-phase-shift})) versus $\Delta_0$ under three different
cases of $\tau$ = $\tau_m$, with $T$ = 1~s and $\Omega_1/\Omega_2$ =
8. The dots show numerical results from Eq.~(\ref{pilot-equation})
and solid curves are analytical results from
Eq.~(\ref{ground-state-phase-shift}). Here, we find a non-vanishing
$\delta\nu$ under all conditions. However, as the ratio of
$\Omega_1/\Omega_2$ increases, we do find both approaches give the
same value of $(\Delta_0)_m$ where clock shift is suppressed, as
indicated in Fig.~\ref{magic-optical-shift}(b). When different free
evolution times ($T$) or pulse durations ($\tau$ = $\tau_m$) are
used, the accumulated phase shift changes, leading to variations in
the dependence of $\delta\nu$ on $\Delta_0$, as shown in the
expanded view of Fig.~\ref{magic-optical-shift}(c) and (d). To
determine the optimum value of $(\Delta_0)_m$ for a practical clock
realization at $\tau$ = 1~s, we can use two different techniques.
First, as shown in Fig.~\ref{magic-optical-shift}(c), extending $T$
reduces the sensitivity of $\delta\nu$ on $\Delta_0$. Hence, the
curves depicting $\delta\nu$ versus $\Delta_0$ for different $T$
rotate around $(\Delta_0)_m$, with no changes in the signal
contrast. In the second approach, as shown in
Fig.~\ref{magic-optical-shift}(d), we find the values of $\Delta_0$
where $\delta\nu$ for $\tau$ = 1~s is the same as that for some
other values of $\tau$ ($<$ 1~s). These values of $\Delta_0$ can be
plotted as a function of $\tau$ and extrapolated to $\tau$ = 0 to
find $(\Delta_0)_m$. However, the signal contrast under smaller
$\tau$ is reduced due to the effect of pulse preparation on
population transfer.

\begin{figure}[!t]
\resizebox{\linewidth}{!}{
\includegraphics[angle=0]{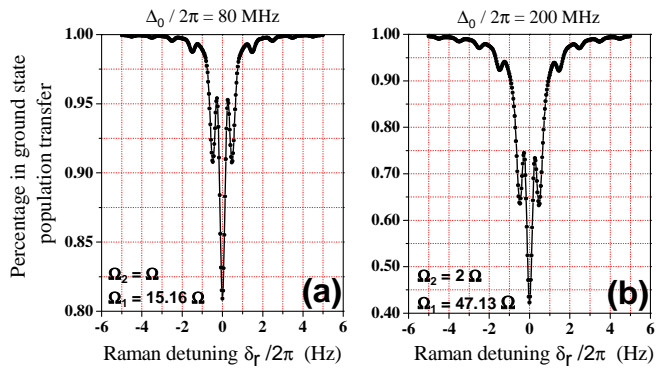}}
\caption{(Color online)Theoretical EIT/Raman lineshapes using
Eq.~(\ref{pulsed-two-photon}) for the clock transition
$^{1}S_{0}\leftrightarrow {^{3}P_{0}}$ and a free evolution time $T
= 1$~s. (a) Population transfer of 20$\%$ under
$\tau_{p}(80$~MHz$)=135$~s and (b) population transfer of nearly
$60\%$ under $\tau_{p}(200$~MHz$)=200$~s. The actual pulse durations
are $\tau = \tau_m = 1$~s.} \label{pulses-ground-state}
\end{figure}

From Eq.~(\ref{pulsed-two-photon}) we find spectral lineshapes and
transition probabilities as a function of the experimental parameter
$\delta_r$, shown in Fig.~\ref{pulses-ground-state}(a). Since
$\tau\ll\tau_{p}(\Delta_{0})$, the two-photon resonance has a
Fourier transform linewidth given by the duration $\tau$ where power
broadening effects have been eliminated. The spectra also exhibit
the typical coherent Ramsey nutations with period $\sim1/2T$ and a
central fringe free from systematic shifts. We have also determined
the sensitivity of $\delta\nu$ to laser intensities ($\Omega_i$) and
detunings, demonstrating that the uncertainty of the optical clock
frequency $<$5~mHz ($\sim10^{-17}$) is achievable by controlling
$\Delta_0$ at the 100~kHz level around $(\Delta_0)_m$. Meanwhile,
$\Omega_i$ fluctuations should be controlled $<$0.5\%. We note that
for a given set of $\tau$ and $\Omega_i$, different values of
$(\Delta_0)_m$ can be found. For example, $(\Delta_0)_m$ = 200 MHz
is another optimum value for larger $\Omega_i$
(Fig.~\ref{pulses-ground-state} (b)). In this case, the signal
contrast is further improved with a population transfer of up to
60\%, leading to enhanced clock stability but also slightly larger
uncertainty.

In summary, our method achieves the $10^{-17}$ accuracy expected for
a ``light-insensitive'' lattice clock with EIT/Raman pulses to
dilute fluctuations of the frequency shift over the free evolution
time $T$. We show that a contrast between $20\%$ to $60\%$
(Fig.~\ref{pulses-ground-state}) could be achieved, also including
realistic lattice decoherence times \cite{Ludlow:2006}. Extensions
are possible to the proposal of \cite{Hong:2005} by replacing the
$^{1}P_{1}$ state with $^{3}P_{1}$, to magnetic field induced
optical transitions \cite{Barber:2006}, for other species like
$^{52}$Cr \cite{Bell:1999}, and for nuclear clock transitions
\cite{Peik:2002}.

We thank J. Dalibard, T. Ido, T. Zelevinsky, and C. Oates for
discussions. This work is supported by ONR, NIST, and NSF. T. Zanon
thanks Observatoire de Paris and D\'{e}l\'{e}gation G\'{e}n\'{e}rale
de l'Armement
for support.\\
\indent *Present address: Laser Cooling and Trapping Group, NIST
Gaithersburg, MD-20899, USA; permanent address: Dipartimento di
Fisica, Universit\`a di Pisa, Italy.

\end{document}